\documentclass[11pt]{article}

\usepackage{acl}

\usepackage{times}
\usepackage{latexsym}
\usepackage{amsmath}
\usepackage{xcolor} 
\usepackage{float}
\usepackage{graphicx}
\usepackage{booktabs}
\usepackage{adjustbox}
\usepackage{geometry}
\usepackage{multirow}
\usepackage{enumitem}
\usepackage[table,dvipsnames]{xcolor}
\usepackage{pifont}
\usepackage{subcaption}
\usepackage[nameinlink,capitalise]{cleveref}
\usepackage{colortbl}
\usepackage[ruled,vlined]{algorithm2e}
\usepackage{booktabs, multirow, siunitx} 
\usepackage{tikz}

\definecolor{LightRed}{RGB}{255,220,220}
\definecolor{MidRed}{RGB}{255,180,180}
\definecolor{LightBlue}{RGB}{220,235,255}
\definecolor{MidBlue}{RGB}{180,210,255}

\newcommand{\tok}[2]{\begingroup
  \setlength{\fboxsep}{1pt}%
  \colorbox{#1!30}{#2}%
\endgroup}
\newcommand{\ourmethod}{\text{LBR} }
\newcommand{\ourmethodName}{\textbf{L}earn \textbf{B}efore \textbf{R}epresent (\textbf{LBR}) }
\newcommand{\ourmethodSft}{\text{IB-GL} }

\newcommand{\mathcolorbox}[2]{\colorbox{#1}{$\displaystyle #2$}}
\usepackage[T1]{fontenc}

\usepackage[utf8]{inputenc}

\usepackage{microtype}

\usepackage{inconsolata}

\usepackage{graphicx}

\title{Learn Before Represent: Bridging Generative and Contrastive Learning for Domain-Specific LLM Embeddings}

\author{
 \textbf{Xiaoyu Liang\textsuperscript{1}},
 \textbf{Yuchen Peng\textsuperscript{1}},
 \textbf{Jiale Luo\textsuperscript{1}},
 \\
 \textbf{Wenhao Wang\textsuperscript{1}},
 \textbf{Haoji Hu\textsuperscript{1}},
 \textbf{Xincheng Zhou\textsuperscript{2}},
\\
\\
 \textsuperscript{1}Zhejiang University,
 \textsuperscript{2}Peking University,
\\
 \small{
   \textbf{Correspondence:} \href{mailto:zhouxincheng@pku.edu.cn}{zhouxincheng@pku.edu.cn}
 }
}

\begin{document}
\maketitle


\begin{abstract}

Large Language Models (LLMs) adapted via contrastive learning excel in general representation learning but struggle in vertical domains like chemistry and law, primarily due to a lack of domain-specific knowledge. This work identifies a core bottleneck: the prevailing ``LLM+CL'' paradigm focuses on semantic alignment but cannot perform knowledge acquisition, leading to failures on specialized terminology. To bridge this gap, we propose Learn Before Represent (LBR), a novel two-stage framework. LBR first injects domain knowledge via an Information Bottleneck-Constrained Generative Learning stage, preserving the LLM's causal attention to maximize knowledge acquisition while compressing semantics. It then performs Generative-Refined Contrastive Learning on the compressed representations for alignment. This approach maintains architectural consistency and resolves the objective conflict between generative and contrastive learning. Extensive experiments on medical, chemistry, and code retrieval tasks show that LBR significantly outperforms strong baselines. Our work establishes a new paradigm for building accurate and robust representations in vertical domains.

\end{abstract}

\section{Introduction}

The rise of Large Language Models (LLMs) has fundamentally reshaped representation learning. Benefiting from their extensive world knowledge and superior language understanding capabilities acquired during large-scale pretraining, LLM-based embedding methods~\cite{llm2vec,qwen3embed} employ contrastive learning (CL) to achieve impressive performance on general benchmarks such as MTEB~\cite{mteb}.

\begin{figure}[!t]
    \includegraphics[width=\columnwidth]{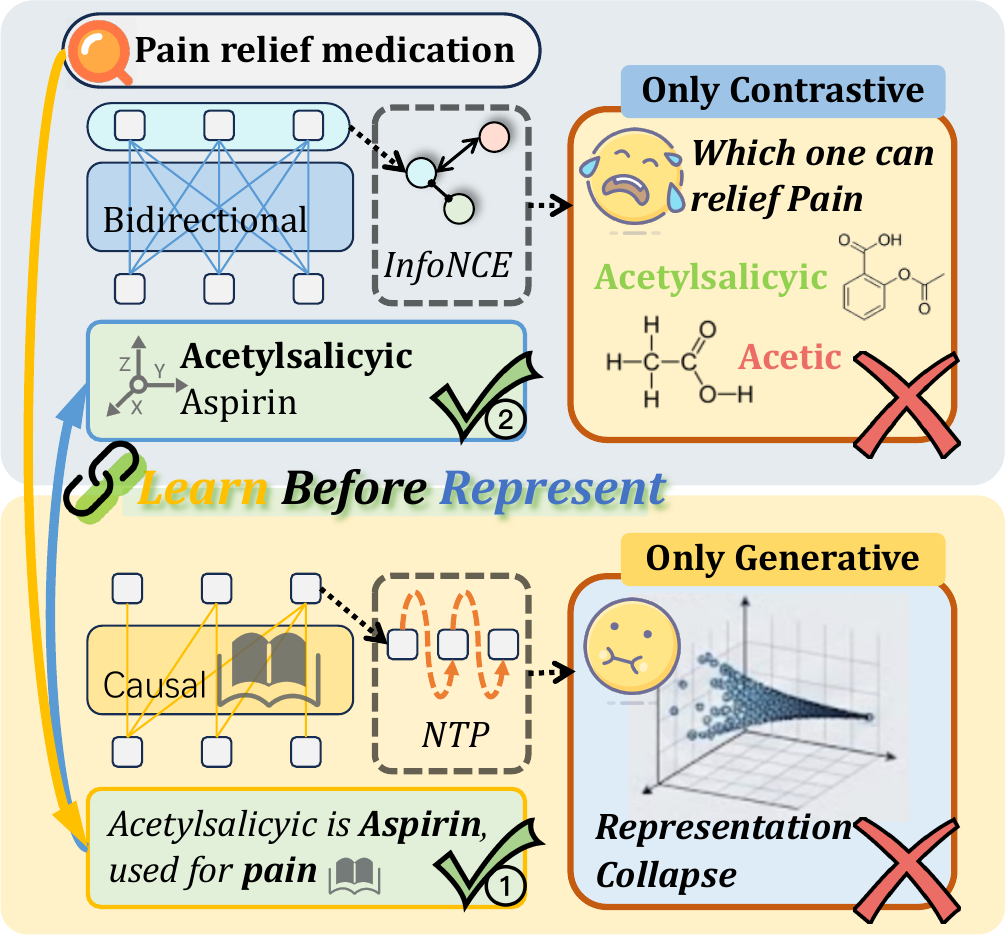}
    \caption{\textbf{Motivation of the Learn Before Represent (LBR) framework.} \mathcolorbox{cyan!15}{(Top)} Only Contrastive (LLM+CL) methods focus on semantic alignment but fail on domain-specific entities (e.g., matching "Acetylsalicylic" to "Aspirin") due to a lack of internal knowledge. \mathcolorbox{yellow!10}{(Bottom)} Only Generative (LLM+GL) methods acquire knowledge via Next-Token Prediction but suffer from representation collapse. We addresses these challenges through a \mathcolorbox{green!10}{\textbf{Learn \ Before \ Represent}} framework.}
    
    \label{fig:first}
\end{figure}

However, this success often fails to translate to vertical domains such as finance and law~\cite{chemteb,finmteb}, where embedding models must handle domain-specific terminology and long-tail entities absent from general pretraining corpora. 
When the LLM itself lacks understanding of domain concepts, the prevalent \textbf{``LLM+CL''} paradigm struggles to construct accurate semantic representations. 
While existing methods attempt to mitigate this issue through hard negative mining~\cite{lei-etal-2023-unsupervised} or constructing vertical domain data~\cite{financialDomain,chemistryDomain}, they overlook a critical limitation: CL inherently focuses on semantic alignment rather than knowledge acquisition.  
Consequently, without first equipping the model with necessary domain knowledge, representation learning in vertical domains remains suboptimal.

To validate this hypothesis, we conduct an in-depth empirical analysis on vertical domain retrieval tasks. Figure~\ref{fig:first}  illustrates a representative case from the chemistry domain: when retrieving pain relievers, a model lacking domain knowledge fails because it does not know that Acetylsalicylic acid is the common pain reliever aspirin. In contrast, a model equipped with domain knowledge achieves precise matching. 
Furthermore, our experimental analysis in Section~\ref{sec:main_results} confirms significant distribution shift between general and vertical domains. 
These findings compel us to move beyond treating LLMs merely as encoders. Instead, we advocate for a \textbf{``LLM+GL+CL''} paradigm: first injecting structured domain knowledge into the LLM via generative learning (GL) (e.g., continued pre-training (PT) or supervised fine-tuning (SFT)), and then refining these representations through CL.

However, realizing this two-stage paradigm poses two fundamental challenges:
\noindent\textbf{Architectural Inconsistency.} 
To enhance embedding quality, existing methods~\cite{llm2vec,nvembed} employ bidirectional attention for contrastive learning. 
However, this architectural shift precludes the model's ability to acquire domain knowledge through autoregressive next-token prediction (NTP). 
Such architectural inconsistency undermines the LLM's potential, preventing learning capabilities and often resulting in catastrophic forgetting or suboptimal alignment~\cite{causal2vec,think&embed}.
\noindent\textbf{Objective Conflict.} 
While GL optimizes token-level cross-entropy for NTP, CL leverages sample-level InfoNCE to shape a global semantic space~\cite{NtpInfonce}. This fundamental mismatch between local generation and global representation frequently exacerbates embedding anisotropy, causing naive integration strategies to fail due to representation collapse~\cite{tsukagoshi2025redundancy,mickus2024isotropy}.


To address these challenges, we propose \textbf{Learn Before Represent (LBR)}, a two-stage framework that unifies knowledge acquisition and representation learning. 
The core idea is to maintain a consistent causal attention architecture, maximizing knowledge learning capability while introducing an \textbf{Information Bottleneck (IB)} to reconcile the potential conflicts between generative and contrastive objectives.
Specifically, in Stage 1 (\textbf{IB-Constrained GL}), we insert bottleneck tokens and mask direct attention from input to target, enabling the model to compress input semantics into the bottleneck tokens and rely solely on these tokens to autoregressively predict the target during GL.
In Stage 2 (\textbf{Generative-Refined CL}), we preserve the causal attention mechanism and leverage the compression capability established in Stage 1 to extract the hidden states of bottleneck tokens as enhanced representations of input sequences, which are further aligned through CL.

The main contributions of this work are summarized as follows:
\begin{itemize}
    \item We propose \ourmethod, a unified two-stage framework that exploits LLMs' knowledge acquisition capability to learn domain-specific knowledge as a foundation for constructing accurate vertical domain representations.

    \item We introduce IB-constrained generative learning, which unifies domain knowledge acquisition and semantic compression within a single objective, effectively bridging generative and contrastive learning while preventing representation collapse.

    \item  We conduct extensive experiments across diverse vertical domains, including chemistry, medical, and code retrieval. Results demonstrate that LBR achieves significant and consistent improvements over strong baselines.

\end{itemize}

\section{Related Work}

\subsection{LLM-based Text Representation Learning}
The paradigm of representation learning is shifting from Encoder-only to Decoder-only LLMs to leverage their superior semantic understanding. 
Pioneering works such as LLM2Vec~\cite{llm2vec} and NV-Embed~\cite{nvembed} propose removing the causal mask during fine-tuning to restore bidirectional attention, effectively attempting to transform a decoder into a globally perceptive encoder. Building on this, 
KaLM-Embedding-V2~\cite{umarvel} and Qwen3-Embedding~\cite{qwen3embed}, have demonstrated that such methods achieve state-of-the-art performance on benchmarks like MTEB~\cite{mteb} via scaling up training data.

\subsection{Preserving Causality: Embedding without Forgetting Generation}
To better leverage the native capabilities of LLMs, recent research has shifted towards enhancing representation capabilities while preserving the Decoder-only architecture.
\noindent\textbf{Reasoning Embeddings.}
{Think\&Embed}~\cite{think&embed} attempts to utilize the Chain-of-Thought (CoT) capability of LLMs to bridge the semantic gap between queries and documents, 
GIRCSE~\cite{gircse} and CCoT~\cite{ccot} propose using ``soft tokens'' as implicit thinking placeholders, maintaining computational depth while reducing overhead.
\noindent\textbf{Reinforcement Learning.}
Methods like GRACE~\cite{grace} reframe contrastive learning as a policy optimization problem within a reinforcement learning framework. Building on this, LREM~\cite{lrem} and TaoSearchEmb~\cite{taosearchemb} incorporate multi-objective optimization strategies to enhance robustness without relying on hard negative mining.
However, these methods primarily target general-purpose reasoning and overlook the critical challenge of domain-specific knowledge acquisition in vertical domains.

\subsection{Representation Enhancement via Information Bottleneck}
The Information Bottleneck (IB) principle offers a theoretical foundation for extracting essential features via compression.
In Computer Vision, Masked Autoencoders~\cite{mae} have proven that ``mask-reconstruction'' is a powerful paradigm for learning robust representations. 
In NLP, methods like UniMAE~\cite{unimae} and RetroMAE~\cite{retromae} implement this via masked auto-encoding. However, they typically rely on asymmetric architectures or auxiliary decoding modules, which complicates optimization and hinders direct adaptation to standard Decoder-only LLMs~\cite{gem,unimae,deng2025following}.
In contrast, we construct an intrinsic information bottleneck by designing a specialized attention mask within the native Decoder-only architecture, eliminating the need for external modules. During training, this mechanism forces the model to compress global semantics into a limited set of bottleneck tokens, achieving seamless integration of generative learning and representation learning within a unified framework.

\section{Method}
\begin{figure*}[!htbp]
    \includegraphics[width=\linewidth]{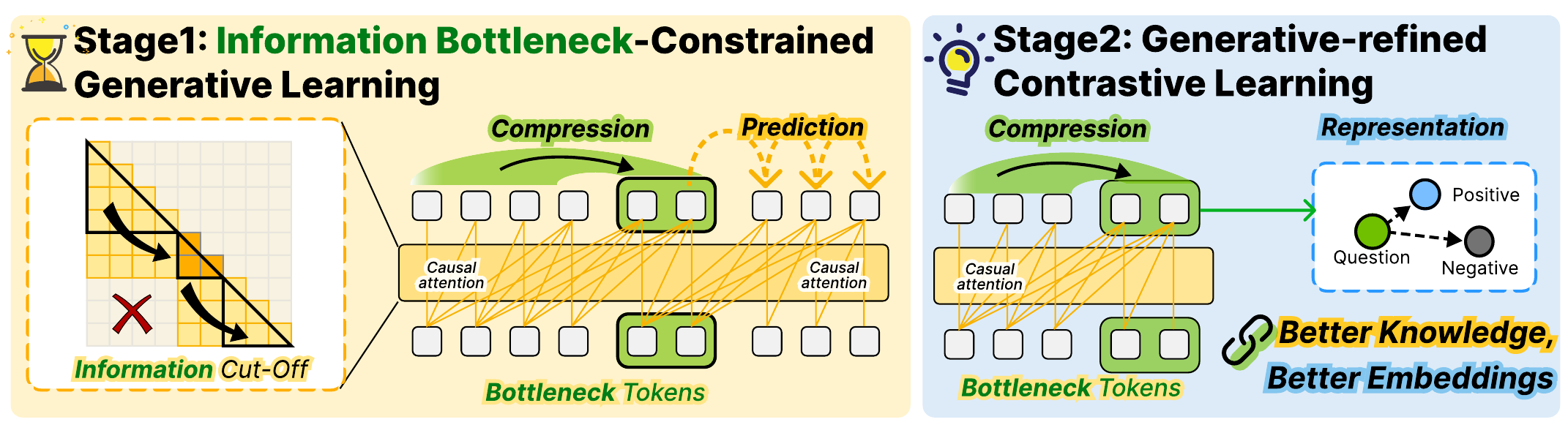}
    \caption{\textbf{Overview of the LBR framework.} In \textbf{Stage 1}, the model performs generative learning under an Information Bottleneck constraint, \mathcolorbox{green!10}{compressing} input semantics into bottleneck tokens and \mathcolorbox{yellow!10}{predicting} the target solely from these tokens. In \textbf{Stage 2}, the bottleneck tokens serve as embeddings, better \mathcolorbox{cyan!15}{representing} the semantics through contrastive learning. Both stages operate under a unified causal attention mechanism to maximize knowledge transfer.}
    \label{fig:overall}
\end{figure*}

We propose \ourmethodName, a two-stage paradigm for adapting LLMs into domain-specific embedding models through domain knowledge infusion (see Figure~\ref{fig:overall}). 
The framework comprises two core components: IB-constrained generative learning (IB-GL) and contrastive learning (CL).
Furthermore, we introduce the \textit{Separation Ratio}, a metric designed to quantify the potential performance gain yielded by GL, thereby guiding the selection of an optimal training trajectory.

\subsection{Preliminaries: The Information Bottleneck Principle}
\label{sec:preliminaries}

The Information Bottleneck principle~\cite{IB,IBCL} posits that an optimal representation should compress the input by discarding irrelevant details, while retaining essential information needed for accurate target prediction.

By introducing bottleneck tokens $Z$ to intercept the flow $X \to Y$, we establish the Markov chain $X \to Z \to Y$ and optimize the following Information Bottleneck objective:
\begin{equation}
    \mathcal{L}_{\text{IB}} = \min \underbrace{\mathcolorbox{green!10}{I(X; Z)}}_{\text{Compression}} - \beta \underbrace{\mathcolorbox{yellow!10}{I(Z; Y)}}_{\text{Prediction}}
\end{equation}
where $X$ and $Y$ represent the source domain knowledge and the generative target, respectively. $I(\cdot; \cdot)$ denotes mutual information, and $\beta$ is a Lagrange multiplier that governs the trade-off between semantic compression (minimizing redundancy) and predictive accuracy (maximizing relevance).

\subsection{Stage 1: IB-Constrained Generative Learning}
While the Information Bottleneck provides a principled framework, directly optimizing mutual information terms is intractable. We therefore instantiate the IB principle through a concrete generative objective.

\paragraph{Minimizing $\mathcolorbox{green!10}{ I(X; Z)}$ via Information Cut-off.}
To enforce the bottleneck constraint and minimize $I(X; Z)$, we employ a specialized attention mask $\mathbf{M}$ that blocks direct information flow from input $X$ to target $Y$. For any target token $i \in Y$, the mask is defined as:
$$\mathbf{M}_{i,j} =
\begin{cases}
1, & \text{if } j \in Z \cup Y_{\le i} \\ 
0, & \text{if } j \in X \quad \textit{(Information Cut-off)}
\end{cases}$$
This architectural constraint establishes a structural bottleneck: by severing the directed path from $X$ to $Y$, the model is compelled to compress all semantic information from $X$ into the limited capacity of bottleneck tokens $Z$.

The feasibility of such extreme compression is supported by recent findings on memory tokens~\cite{memorytokens}, which demonstrate that LLMs can distill hundreds of tokens into a single embedding while maintaining full reconstructability. Building on this insight, we parameterize the bottleneck strength via the compression ratio $R = |X|/|Z|$, defined as the ratio of input length to bottleneck token count. A critical question then arises: what compression ratio yields optimal retrieval representations? Intuitively, excessive compression (large $R$) risks information loss, while insufficient compression (small $R$) retains noise and redundancy. We empirically investigate this trade-off in Section~\ref{sec:compression_ratio}, analyzing how varying $R$ affects downstream retrieval performance.

\paragraph{Maximizing $\mathcolorbox{yellow!10}{ I(Z; Y)}$ via Generative Loss.}
With the information flow constrained to pass through $Z$, we maximize $I(Z; Y)$ to ensure that $Z$ retains sufficient information for target reconstruction. Specifically, we employ a standard autoregressive next-token prediction (NTP) objective:
\begin{equation}
    \mathcal{L}_{\text{gen}} = - \sum_{t=1}^{|Y|} \log P(Y_t \mid Z, Y_{<t}; \theta)
\end{equation}
Minimizing $\mathcal{L}_{\text{gen}}$ ensures that the bottleneck tokens $Z$ preserve all information necessary to accurately predict the target sequence $Y$.

Depending on data availability, this phase has two variants, both designed to instill domain knowledge.
When labeled query-answer pairs are available, we adopt SFT-style GL with $X = Q$ and $Y = A$, guiding the model to compress query semantics into $Z$ for answer generation.
When only unlabeled domain corpora are available, we employ PT-style generative learning via self-supervised objectives: either passage reconstruction ($X = Y = D$) or prefix-suffix prediction ($X = D_{:k}, Y = D_{k:}$). This variant enables unsupervised knowledge acquisition at the cost of increased computation (see Section~\ref{app:pt_analysis} for detailed analysis).

\paragraph{Information-Theoretic Interpretation.}
Our approach instantiates the IB principle through a dual mechanism. 
The attention mask $\mathbf{M}$ and compression ratio $R$ impose a hard capacity constraint, implicitly minimizing $I(X; Z)$ by forcing the model to discard noise and redundancy.
Concurrently, the generative loss $\mathcal{L}_{\text{gen}}$ maximizes $I(Z; Y)$, ensuring that $Z$ retains features critical for domain-specific reconstruction.
Crucially, beyond merely acquiring domain knowledge, this generative stage trains the model to effectively aggregate and compress semantics into $Z$.

\subsection{Stage 2: Generative-Refined Contrastive Learning}
\label{sec:stage2}
Building on the semantic compression capability established in Stage 1, Stage 2 aligns representations through \textbf{Generative-Refined Contrastive Learning (GR-CL)}.
We preserve the native causal attention mechanism to ensure architectural consistency with Stage 1.

Formally, given the input structure $[X; Z]$ where $X$ denotes the domain sequence and $Z$ the bottleneck tokens, we extract the hidden state of the final token in $Z$ as the sequence representation $\mathbf{v}$.
We align query-passage pairs via the InfoNCE loss:
\begin{equation}
\mathcal{L}_{\text{contrast}} = -\log\frac{e^{s(q,p^+)/\tau}}{e^{s(q,p^+)/\tau} + \sum_{p^-} e^{s(q,p^-)/\tau}}
\end{equation}
where $s(q,p) = \text{sim}(\mathbf{v}_q, \mathbf{v}_p)$, $\tau$ is the temperature, and the sum runs over in-batch negatives.

Unlike methods that rely on auxiliary decoders or structural changes~\cite{unimae,deng2025following,su2025training}, our approach preserves the LLM's native architecture and generative capacity for domain knowledge acquisition. 
Preserving the causal attention mechanism yields two key benefits: (1) \textbf{Consistency}: It ensures a seamless transition from Stage 1, enabling the model to fully leverage the acquired domain knowledge and compression abilities; (2) \textbf{Extensibility}: It retains the LLM's inherent strengths, including instruction-following capabilities while paving the way for future inference-time reasoning, such as Chain-of-Thought~\cite{gircse}.

\section{Experiments}

\subsection{Experimental Settings}  
\paragraph{Datasets.}
To evaluate our method across specialized domains, we curate datasets from multiple vertical domains, including medicine, finance, and code. For each domain, we extract 150k samples for the GL stage and 50k samples for the CL stage. 
To ensure consistent input formatting, we standardize all raw data—including question-answer, query-document, and multiple-choice formats—into a unified QA format suitable for subsequent GL and supervised CL. The dataset construction details are provided in Appendix~\ref{app:appendixdata}.

\paragraph{Evaluation Protocol.}
We evaluate retrieval performance using the MTEB~\cite{mteb} benchmark framework. For each vertical domain, we reserve an additional 20k samples as evaluation sets for retrieval tasks. These evaluation sets are strictly isolated from training data, ensuring no data leakage.

\paragraph{Baselines.}
We compare our method against several categories of baselines on specialized domain datasets:
\noindent\textbf{Domain-Adapted LLMs.} This category includes pretrained language models continually trained on domain-specific corpora to enhance domain expertise without task-specific supervision, such as ChemLLM~\cite{yao2024lawyer} for chemistry and HuaTuo~\cite{huatuo} for medical domain.
\noindent\textbf{Contrastive Learning-Based Embedders.} These methods optimize sentence embeddings through contrastive learning, including LLM2VEC~\cite{llm2vec}, GTE~\cite{gte},and BGE~\cite{xiao2024c}.

\subsection{Main Results in Domain}
\label{sec:main_results}
\begin{table*}[t]
\small 
\setlength{\tabcolsep}{2.5pt} 

\setlength{\heavyrulewidth}{0.12em}
\setlength{\lightrulewidth}{0.05em}
\setlength{\cmidrulewidth}{0.07em}
\setlength{\aboverulesep}{0.05pt}
\setlength{\belowrulesep}{0.05pt}

\resizebox{\textwidth}{!}{
\definecolor{MidBlue}{RGB}{135,206,250}
\definecolor{LightBlue}{RGB}{176,224,230}
\definecolor{LightRed}{RGB}{255,182,193}

\begin{tabular}{l l c |c c c c c c c}
    \toprule
    \multirow{2.5}{*}{\textbf{Models}} & \multirow{2.5}{*}{\textbf{Backbone}} & \multirow{2.5}{*}{\textbf{Attn.}} & 
    \multicolumn{2}{c}{\textbf{Chemistry}} & \multicolumn{2}{c}{\textbf{Medical}} & \multicolumn{2}{c}{\textbf{Code}} & \multirow{2.5}{*}{\textbf{Avg. (Rank)}} \\
    
    \cmidrule(lr){4-5} \cmidrule(lr){6-7} \cmidrule(lr){8-9}
     & & & R@10 & N@10 & R@10 & N@10 & R@10 & N@10 &  \\
    \midrule
    
    \multicolumn{10}{l}{\textit{\textbf{Pre-trained LLMs} \tok{yellow}{(LLM+GL)}}} \\ 
    Qwen2 & Qwen2-1.5B & Causal & 0.100 & 0.078 & 0.410 & 0.335 & 0.151 & 0.108 & 20.0 (10) \\
    Qwen2.5 & Qwen2.5-1.5B & Causal & 0.146 & 0.115 & 0.353 & 0.286 & 0.116 & 0.081 & 18.3 (11) \\
    ChemLLM & InternLM-2B & Causal & 0.237 & 0.168 & 0.034 & 0.020 & 0.116 & 0.082 & 11.0 (12) \\
    HuaTuo & LLaMA-7B & Causal & 0.201 & 0.188 & 0.346 & 0.283 & 0.130 & 0.101 & 31.2 (9) \\
    
    \specialrule{\lightrulewidth}{2pt}{1.2pt} 
    \specialrule{\lightrulewidth}{0pt}{0pt}
    
    \multicolumn{10}{l}{\textit{\textbf{Contrastive Learning-based Methods} \tok{yellow}{(LLM+CL)}}} \\ 
    
    BGE & XLMR-Large & Bi. & 0.689 & 0.557 & 0.947 & 0.871 & 0.882 & 0.754 & 78.3 (5) \\
    GTE & Qwen2-1.5B & Bi. & 0.766 & 0.677 & 0.958 & 0.867 & 0.360 & 0.286 & 65.2 (6) \\
    LLM2Vec & SLLaMA-1.3B & Bi. & 0.712 & 0.615 & 0.961 & 0.890 & 0.850 & 0.729 & 79.3 (4) \\

    \specialrule{\lightrulewidth}{2pt}{1.2pt} 
    \specialrule{\lightrulewidth}{0pt}{0pt}
    
    \multicolumn{10}{l}{\textit{\textbf{Contrastive Learning on Domain LLMs} \tok{yellow}{(LLM+GL+CL)}}} \\ 
    
    \texttt{SFT+CL} & Qwen2-1.5B & Causal & 0.449 & 0.405 & 0.732 & 0.625 & 0.888 & 0.764 & 64.4 (7) \\
    \texttt{SFT+CL} & Qwen2.5-1.5B & Causal & 0.436 & 0.391 & 0.664 & 0.558 & 0.855 & 0.732 & 60.6 (8) \\

    \specialrule{\lightrulewidth}{2pt}{1.2pt} 
    \specialrule{\lightrulewidth}{0pt}{0pt}
    
    \multicolumn{10}{l}{\textit{\textbf{Ours Method} \tok{yellow}{(LLM+IB-GL+CL)}}} \\ 
    \texttt{LBR} & Llama3.2-1B & Causal & 0.722 & 0.619 & 0.976 & 0.897 & 0.979 & 0.879 & 84.5 (3) \\
    \texttt{LBR} & Qwen2-1.5B & Causal & \textbf{0.802} & \textbf{0.727} & 0.978 & 0.899 & 0.978 & 0.882 & 87.8 (2) \\
    \rowcolor{cyan!15}\texttt{LBR} & Qwen2.5-1.5B & Causal & 0.797 & 0.723 & \textbf{0.979} & \textbf{0.906} & \textbf{0.980} & \textbf{0.887} & \textbf{87.9 (1) }\\

    \bottomrule
\end{tabular}
}
\caption{Performance comparison on terminology understanding tasks across three vertical domains (Chemistry, Medical, Code). 
\textbf{R@10} and \textbf{N@10} represent Recall@10 and NDCG@10, respectively.
\textbf{Attn.} denotes the model's attention architecture (\textit{Causal} for unidirectional or \textit{Bi.} for bidirectional).
\tok{yellow}{Highlighted tags} categorize the training paradigms: 
\textbf{LLM+GL}: Standard Generative Loss (SFT) baselines; 
\textbf{LLM+CL}: Contrastive Learning-based methods; 
\textbf{LLM+GL+CL}: naively combining GL and CL; 
\textbf{LLM+IB-GL+CL}: Our proposed method incorporating Information Bottleneck. 
\textbf{Bold} indicates the best performance in each column.
\textbf{Score (Rank)} summarizes the overall metric and relative ranking.}

\label{tab:main}
\end{table*}

Table~\ref{tab:main} compares performance across Chemistry, Medical, and Code domains.

\texttt{LBR} consistently achieves the best results. Specifically, \texttt{LBR} (Qwen2.5-1.5B) significantly outperforms the strongest baseline LLM2Vec (Avg Score: 87.9 vs. 79.3). Notably, our smaller llama3.2-1B model surpasses larger baselines like XLMR-Large and Qwen2-1.5B, demonstrating superior retrieval efficiency.

\noindent\textbf{Impact of Bottleneck Constraint.}
While the naive \texttt{SFT+CL} lags behind contrastive baselines due to representation collapse, \texttt{LBR} utilizes the Information Bottleneck to enforce semantic compression. This constraint yields dramatic improvements, boosting R@10 by \textbf{+36.6\%} (0.436 to 0.802) in Chemistry compared to \texttt{SFT+CL}.

\noindent\textbf{Causal vs. Bidirectional.}
Unlike LLM2Vec which relies on bidirectional adaptation, \texttt{LBR} preserves native causal attention yet outperforms bidirectional models (e.g., +9.0\% R@10 in Chemistry). This indicates that with effective semantic compression, causal architectures can serve as superior embedding models without structural modifications.

\subsection{Representation and Generation}
\label{sec:dual_capability}

\begin{table}[t]
\centering
\small  
\setlength{\tabcolsep}{10pt} 

\begin{tabular}{l|cc|cc}
    \toprule
    \multirow{2}{*}{\textbf{Method}} & \multicolumn{2}{c|}{\textbf{Retrieval}} & \multicolumn{2}{c}{\textbf{Generation}} \\
    & R@10 & N@10 & B-4 & R-L \\
    \midrule
    Base & 41.03 & 26.68 & 4.42 & 23.71 \\
    GL & 54.91 & 45.98 & \textbf{14.32} & \textbf{35.79} \\
    CL & \underline{75.59} & \underline{64.35} & -- & -- \\
    \rowcolor{LightBlue!66} 
    IB-GL & \textbf{90.33} & \textbf{80.60} & \underline{13.98} & \underline{35.58} \\
    \bottomrule
\end{tabular}

\caption{\textbf{Representation and generation capability on the medical domain.} We report Recall@10 (R@10) and NDCG@10 (N@10) for retrieval tasks, alongside BLEU-4 (B-4) and ROUGE-L (R-L) scores for generation tasks. The \textbf{best} performance is highlighted in bold, and the \underline{second best} is underlined.}
\label{tab:dual_capability}
\end{table}



To validate that our IB-GL paradigm effectively acquires domain knowledge while enhancing representation capabilities, we evaluated four model variants: \textbf{Base} (pretrained Qwen2-1.5B), \textbf{SFT} (standard supervised fine-tuning), \textbf{CL} (contrastive learning only), and \textbf{IB-GL} (ours).

As shown in Table~\ref{tab:dual_capability}, \textbf{SFT} achieves the highest generation scores (14.32 B-4), confirming its effectiveness in internalizing domain knowledge. However, while SFT improves retrieval over the base model (54.91 vs. 41.03 R@10), it significantly lags behind contrastive methods, indicating that next-token prediction alone is insufficient for learning robust semantic representations.
\textbf{CL} substantially boosts retrieval (75.59 R@10) but suffers from catastrophic forgetting of generative capability (indicated by --).

In contrast, \textbf{IB-GL} achieves the best of all methods. It delivers state-of-the-art retrieval performance (\textbf{90.33} R@10), outperforming CL by a large margin ($\uparrow$14.74). Crucially, it preserves the generative capability, maintaining scores comparable to SFT (13.98 vs. 14.32 B-4). This result validates our core hypothesis: IB-GL successfully unifies \textbf{domain knowledge acquisition} with \textbf{semantic compression}, providing a superior representation without sacrificing the LLM's generative nature.

\subsection{Analysis of different stages in LBR}

\definecolor{LemonChiffon}{RGB}{255, 240, 205}
\definecolor{lavender}{RGB}{230, 230, 250}

\begin{table*}[ht]
\vspace{-3mm}
\footnotesize 
\centering

\resizebox{1.0\textwidth}{!}{%
    
    \begin{tabular}{l|l r| l r}
        \toprule
        Stages & \multicolumn{2}{c|}{\textcolor[RGB]{172, 172, 234}{Medical}} & \multicolumn{2}{c}{\textcolor[RGB]{241, 205, 122}{Code}} \\
        \midrule
         
        {LLM} & 
        \begin{tikzpicture}\filldraw[lavender] (0,-0.1) rectangle (0.934,0.12);\end{tikzpicture} & 33.52 & 
        \begin{tikzpicture}\filldraw[LemonChiffon] (0,-0.1) rectangle (0.377,0.12);\end{tikzpicture} & 10.78 \\
        \midrule
         
        LLM+GL & 
        \begin{tikzpicture}\filldraw[lavender] (0,-0.1) rectangle (1.609,0.12);\end{tikzpicture} & 45.98 & 
        \begin{tikzpicture}\filldraw[LemonChiffon] (0,-0.1) rectangle (1.425,0.12);\end{tikzpicture} & 40.71 \\
         
        LLM+IB-GL & 
        \begin{tikzpicture}\filldraw[lavender] (0,-0.1) rectangle (2.821,0.12);\end{tikzpicture} & 80.60 & 
        \begin{tikzpicture}\filldraw[LemonChiffon] (0,-0.1) rectangle (2.566,0.12);\end{tikzpicture} & 73.31 \\
        \midrule
        LLM+CL & 
        \begin{tikzpicture}\filldraw[lavender] (0,-0.1) rectangle (2.252,0.12);\end{tikzpicture} & 64.35 & 
        \begin{tikzpicture}\filldraw[LemonChiffon] (0,-0.1) rectangle (2.692,0.12);\end{tikzpicture} & 76.91 \\

        LLM+GL+CL & 
        \begin{tikzpicture}\filldraw[lavender] (0,-0.1) rectangle (2.186,0.12);\end{tikzpicture} & 62.45 & 
        \begin{tikzpicture}\filldraw[LemonChiffon] (0,-0.1) rectangle (2.673,0.12);\end{tikzpicture} & 76.36 \\
         
        LLM+IB-GL+CL & 
        \begin{tikzpicture}\filldraw[lavender] (0,-0.1) rectangle (3.109,0.12);\end{tikzpicture} & 89.85 & 
        \begin{tikzpicture}\filldraw[LemonChiffon] (0,-0.1) rectangle (3.088,0.12);\end{tikzpicture} & 88.22 \\
         
        \bottomrule
    \end{tabular}%
}
\caption{Comparison of Retrieval Performance Across Different Stages in the Medical and Code Domains. We use Qwen2-1.5B  as the backbone model to assess the incremental improvements of subsequent training stages.}
\label{tab:ablstages}
\end{table*}
To thoroughly analyze the specific contributions of each stages, we conduct ablation studies in Table~\ref{tab:ablstages}.

\paragraph{Impact of Standard GL.} Introducing domain-specific SFT yields only marginal improvements on a subset of tasks. While this partially validates the utility of domain knowledge for retrieval, the limited gains reveal that GL alone is insufficient. 

\paragraph{Impact of IB-GL (Ours).} In contrast, our IB-constrained GL achieves substantial performance gains over standard GL. This demonstrates that training the LLM to compress semantics into bottleneck tokens is crucial for effective representation learning while preserving generative capability.

\paragraph{Synergy with CL.} 
When further combined with CL, our method achieves the best overall performance, validating the LBR paradigm. 
Notably, the ``GL + CL'' underperforms CL alone in the medical domain. This confirms our hypothesis articulated earlier: standard GL's token-level optimization conflicts with CL's sample-level objective, leading to representation collapse. Our IB-constrained approach resolves this conflict by enforcing semantic compression during the generative stage.

\section{Ablation Studies}

\subsection{Causal vs. Bidirectional Attention}
\label{sec:ablation_attention}
\begin{table}[t]
\centering
\small
\setlength{\tabcolsep}{3pt} 

\resizebox{0.95\linewidth}{!}{%
    \begin{tabular}{lc|@{\hspace{12pt}}c@{\hspace{12pt}}c}
    \toprule
    \multirow{2}{*}{\textbf{Method}} & \multirow{2}{*}{\textbf{Base Attn.}} & \multicolumn{2}{c}{\textbf{Attention Mechanism in CL}} \\ 
    \cmidrule(l{12pt}r){3-4} 
     & & \textbf{Causal (Ours)} & \textbf{Bidirectional} \\
    \midrule
    GL & Causal & 54.91 & 55.04 \\
    \textbf{IB-GL (Ours)} & Causal & \textbf{75.59} & 64.35 \\
    \bottomrule
    \end{tabular}%
}
\caption{\textbf{Impact of Attention Mechanisms during Contrastive Learning.} 
We compare the retrieval performance using Causal (Unidirectional) versus Bidirectional attention masks. All base models utilize a Causal attention architecture. }
\label{tab:AttentioninAlignment}
\end{table}

Recent LLM-based embedding methods~\cite{llm2vec} typically remove the causal mask during fine-tuning to enable bidirectional attention. However, we argue this disrupts the autoregressive nature of LLMs. We compare two CL variants after IB-GL training: one with \textbf{Bidirectional Attention} and one with \textbf{Causal Attention}.

As shown in table~\ref{tab:AttentioninAlignment}, IB-GL+Causal CL outperforms IB-GL+Bidirectional CL, contradicting the intuition that bidirectional context yields better representations.
We attribute this to two key factors:
\textit{(1) Alignment with IB-GL.} IB-GL trains the model to compress information into bottleneck tokens autoregressively. Switching to bidirectional attention creates a distribution shift that breaks this learned compression ability.
\textit{(2) Knowledge Activation.} Causal attention forces the model to use its internal knowledge for prediction, where as bidirectional attention enables shallow pattern matching across all tokens, bypassing deeper reasoning capabilities.

\noindent\textbf{Broader Impact.}
Maintaining causal attention preserves LLMs' reasoning potential, making our approach compatible with emerging CoT-enhanced embedding methods~\cite{think&embed,gircse} that leverage chain-of-thought capabilities for representation learning.

\subsection{Bottleneck Compression Ratio}
\label{sec:compression_ratio}
We evaluate five compression ratios $R \in \{10, 20, 100, 500, 1000\}$. 
The bottleneck strength is controlled by the compression ratio $R = L_{\text{input}} / N_{\text{tokens}}$, where $L_{\text{input}}$ denotes the input length and $N_{\text{tokens}}$ denotes the number of bottleneck tokens. 
For efficiency, we use Qwen2-1.5B and train each configuration for 1,000 steps using Stage 1 only, then directly evaluate on retrieval tasks to isolate compression effects.


\begin{table}[t]
\centering
\small
\begin{tabular}{l|cc}
    \toprule
    \textbf{Compression Ratio} & \textbf{NDCG@10} & \textbf{Recall@10} \\
    \midrule
    $R=10$   & 8.17  & 10.65 \\
    $R=20$   & 10.84 & 13.26 \\
    $R=50$   & 17.14 & 20.66 \\
    $R=100$  & 18.75 & 21.98 \\
    \rowcolor{LightBlue!66}$R=500$  & \textbf{27.31} & \textbf{30.09} \\ 
    $R=1000$ & 20.97 & 23.57 \\
    \bottomrule
\end{tabular}
\caption{\textbf{Impact of compression ratio on retrieval performance.} We evaluate six compression ratios 
by stratifying training samples by length and assigning different numbers of bottleneck tokens accordingly.}
\label{tab:compression_ratio}
\end{table}

Table~\ref{tab:compression_ratio} reveals that mild compression ($R=10, 20$) provides insufficient bottleneck pressure, while extreme compression ($R=1000$) causes information loss. Optimal performance occurs at $R=500$, with stable performance across $R \in [200, 500]$, demonstrating robustness to moderate variations. 
We recommend $R=500$ as the default setting, with adjustments to $R \in [200, 400]$ for information-dense domains (e.g., medical) or $R \in [500, 800]$ for high-redundancy domains.

\subsection{Learning Efficiency and Data Allocation}
\label{sec:efficiency}
To demonstrate that performance gains stem from our method rather than increased data volume, we conduct a data allocation study with a fixed budget of 100k samples.

\begin{figure}[t]
    \centering
    \includegraphics[width=0.48\textwidth]{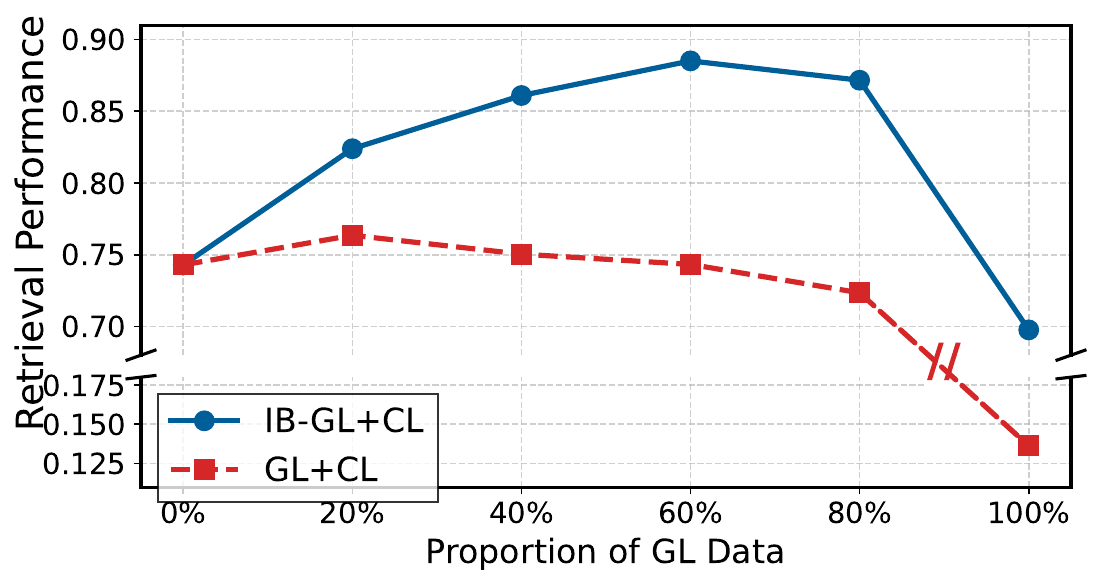}
    \caption{\textbf{Impact of GL data allocation.} We fix the total training budget at 100k samples and vary the GL data ratio ($r_{\text{learn}}$) from 0\% to 100\% in Stage 1, where $r_{\text{learn}}=0\%$ represents pure contrastive learning and $r_{\text{learn}}=100\%$ represents pure generative learning. Both our IB-constrained method and standard GL are evaluated under identical data allocation settings.}
    \label{fig:learningefficiency}
\end{figure}

Figure~\ref{fig:learningefficiency} reveals inverted U-shaped curves for both methods. As $r_{\text{stage1}}$ increases, performance initially improves, confirming that domain knowledge injection is beneficial. However, excessive GL data allocation degrades performance, indicating that generative and contrastive learning are complementary—neither alone suffices.

Notably, our IB-constrained approach consistently outperforms standard generative learning across all allocation ratios. 
The steeper ascending slope indicates more efficient knowledge utilization for representation learning, validating that explicit semantic compression better leverages acquired knowledge for retrieval.

\subsection{Training Efficiency Analysis}
\label{app:pt_analysis}

We analyze the computational overhead of IB-GL under both SFT and PT paradigms, comparing against their respective baselines.

\paragraph{IB-GL with SFT.}

Our IB-GL approach under the SFT paradigm incurs minimal computational overhead compared to standard SFT, with the training time increasing by only 3\%. This efficiency stems from our architectural design: we introduce no additional modules, merely inserting a small number of bottleneck tokens and modifying the attention mask through parallelized matrix operations. Consequently, the computational complexity remains nearly identical to standard SFT.

\paragraph{IB-GL with PT.}

The PT paradigm exhibits higher computational costs due to the reconstruction task. Since the input sequence must be reconstructed, the sequence length doubles, 
resulting in approximately $4\times$ training time due to the quadratic attention complexity.

\paragraph{SFT vs. PT.}Table~\ref{tab:sft_vs_pt} compares the retrieval performance and training efficiency of both paradigms. SFT achieves slightly better retrieval performance than PT, as the carefully curated query-answer pairs enable more efficient learning. The PT paradigm, while less efficient, provides a viable alternative when supervised data is unavailable, trading computational cost for data scalability.

\begin{table}[t]
\centering
\small
\begin{tabular}{lccc}
\toprule
\textbf{Method} & \textbf{Retrieval} & \textbf{Training Time} \\
\midrule
Standard SFT & 45.98 & 1.0$\times$ \\
IB-GL (SFT) & 80.60 & 1.03$\times$ \\
\midrule
Standard PT & 43.25 & 1.0$\times$  \\
IB-GL (PT) &  81.36 & 3.9$\times$  \\
\bottomrule
\end{tabular}
\caption{Performance and efficiency comparison between SFT and PT paradigms. IB-GL (SFT) achieves superior retrieval performance with negligible overhead, while IB-GL (PT) offers a scalable solution at higher computational cost.}
\label{tab:sft_vs_pt}
\end{table}

In practice, we recommend SFT when labeled data is available for its superior efficiency. PT serves as an alternative for scenarios with abundant unlabeled corpora but limited supervision.

\section{Conclusion}
In this work, we introduced \ourmethod, a unified two-stage framework that resolves the fundamental conflicts between generative and contrastive learning in vertical domain adaptation. Our core insight posits that domain knowledge ingestion must precede representation alignment. 

To realize this, \ourmethodSft\ leverages an Information Bottleneck mechanism to enforce ``representation-friendly'' generative learning. By compressing global semantics into bottleneck tokens without altering the native architecture, our approach enables effective knowledge injection while optimally initializing the model for subsequent contrastive alignment. Extensive experiments across medical, chemistry, and code domains validate the superiority of our paradigm over standard methods. 

While this work establishes the efficacy of a sequential two-stage framework, exploring the joint optimization of generative and contrastive objectives remains a promising avenue. We leave this exploration as an intriguing direction for future research.

\section{Limitations}
While our proposed LBR framework demonstrates substantial improvements in vertical domain representation learning, several limitations remain that present opportunities for future work.

First, regarding domain scope, our current evaluation primarily focuses on entity-dense domains such as medicine, chemistry, and code. Although the Information Bottleneck principle is theoretically domain-agnostic, empirical validation on additional specialized fields—particularly those requiring rigorous logic like law and finance—would further strengthen claims of generalizability.

Second, regarding cognitive capability, our framework currently validates the injection of factual domain knowledge (e.g., terminology and entity relationships). However, high-level reasoning and complex logical deduction, which are critical in domains like legal reasoning and financial risk assessment, remain challenging to acquire through the current generative objective alone. A promising direction is to extend LBR by incorporating reasoning-specific supervision (e.g., Chain-of-Thought) during the generative stage, thereby leveraging the preserved causal architecture to handle more complex professional tasks.

Finally, regarding architectural flexibility, the current implementation relies on fixed-length bottleneck tokens to compress semantic information. While effective, this design may be suboptimal for inputs with highly variable information density, potentially leading to under-compression of complex semantics or over-compression of simple ones. Future work could investigate adaptive and hierarchical bottleneck mechanisms where compression intensity is dynamically learned based on input complexity, paving the way for more efficient and expressive representations.



\bibliography{custom.bib}

\end{document}